\documentstyle[12pt]{article}
\begin{document}
\title{{\bf QUANTUM MECHANICS \\ AS A SIMPLE GENERALIZATION
 \\ OF CLASSICAL MECHANICS}
\thanks{Alberta-Thy-06-02, quant-ph/0204015}}
\author{
Don N. Page
\thanks{Internet address:
don@phys.ualberta.ca}
\\
CIAR Cosmology Program, Institute for Theoretical Physics\\
Department of Physics, University of Alberta\\
Edmonton, Alberta, Canada T6G 2J1
}
\date{(2002 April 3)}

\maketitle
\large
\begin{abstract}
\baselineskip 16 pt

A motivation is given for expressing classical mechanics
in terms of diagonal projection matrices and diagonal density matrices.
Then quantum mechanics is seen to be a simple
generalization in which one replaces the diagonal real matrices
with suitable Hermitian matrices.

\end{abstract}
\normalsize
\baselineskip 16 pt

\section{Perception Probabilities in a Classical Model}

\hspace{.25in}The meeting near Princeton, March 15-18, 2002, ``Science
\& Ultimate Reality: Celebrating the Vision of John Archibald Wheeler
and Taking It Forward into a New Century of Discovery,''
reminded us of Wheeler's key question, ``Why the quantum?''
Two answers given at the meeting,
expressing rather opposite motivations,
were that nature is quantum because information is quantized
\cite{Zei},
and that nature is quantized because it abhors ``damn classical jumps''
\cite{Har}.

	Here a more modest approach is taken,
simply showing how one way of formulating classical mechanics
allows quantum mechanics to be obtained by a very simple generalization.

	This approach introduces probabilities
by postulating a perceiver who has access only
to part of the information in the dynamical system,
so that what it perceives can be interpreted probabilistically
in a certain sense, even though the complete description
is completely deterministic.
The rules for the probabilities can then be naturally formulated
in a way that allows a simple generalization of the classical
system to get a quantum system for the probability rules.

	For simplicity, consider a classical system with a time $t$
and a discrete set of $n$ states, ${i=1,\ldots,n}$.
Suppose that the state is a unique function of the time, so $i=i(t)$,
and that the evolution is periodic, with period $T$,
endlessly cycling through each of the $n$ states in turn,
so that in each period the state $i$ occurs for a time duration $f_iT$.
Then each state occurs for a fraction of the time given by ${f_i}$,
assuming that one takes a total time that is an integer multiple of $T$
or else takes the limit of an infinite total time.
This is then taken to be a model of the underlying classical system,
which is completely deterministic ($i(t)$ uniquely given).

	Now suppose there is also a perceiver
that has no direct perception of the time $t$
but can only perceive the state $i$ without
knowing what $t$ is, getting the corresponding perception $p_i$.
For simplicity, assume that the perception of each distinct state
is distinct, so $p_i=p_j$ if and only if $i=j$.
In the full theory, with $i=i(t)$, we thus have $p_i = p_i(t)$,
a unique function of $t$.
But because the perceiver has no direct perception of $t$,
for the perceiver only the perceptions $p_i$
and their time fractions $f_i$ are relevant.
The perceiver does not even have a direct perception of any of the time
fractions, but if one assumes that the measure for each perception
is proportional to the relative time spent on it,
over an infinite total time (or over any integer multiple of $T$),
the normalized measure for each perception is simply the fraction $f_i$.

	(A perceiver is a postulated being
somewhat analogous to an observer,
except that it is not part of the dynamical physical system
and has no effect upon it, which is allowed even in quantum mechanics.
For example, a perceiver might be a nonphysical mind
in a dualist view in which the mind and its perceptions
are epiphenomena that supervene upon the dynamical physical system
without having any action back upon it.
Such a perceiver is not really necessary for my argument,
but it provides a convenient way to divide properties of
the dynamical system into three classes,
those that are putatively directly perceptible, like the state $i$,
those that are not directly perceptible but still are relevant
for the perceiver, like the fraction $f_i$,
and those that are neither perceptible nor relevant to the perceiver,
like the time $t$.)

	Although all of this is completely deterministic,
for the perceiver it can be interpreted
as if it has the perception $p_i$ with probability $f_i$.
To put it another way, though the dynamical system and the corresponding
perceptions occur deterministically, with no uncertainty
with respect to a global viewpoint in which $i(t)$ and hence $p_i(t)$
are known, if one imagines randomly sampling the perceptions
with a sampling measure being proportional to the time that
the perceptions occur, one will get that the perception $p_i$
occurs with probability $f_i$.
Thus probabilities enter this deterministic system
only with a random sampling.
If the sampling is purely hypothetical,
then so is the interpretation of the fractions $f_i$ as probabilities,
but interpreting these time fractions as probabilities
can be highly useful and so will be done here.

	One might be interested not merely in the probabilities
of the individual perceptions $p_i$, but also in various sets
of them, say $S$.  The probability for each set of perceptions,
say $f(S)$, can of course be formed in the usual way,
by adding up the probabilities of the individual perceptions
in the set,
 \begin{equation}
 f(S) = {\bf\chi}\cdot{\bf f} \equiv \sum_{i=1}^n \chi_i f_i,
 \label{1}
 \end{equation}
where $\chi_i$ is the characteristic function that is $\chi_i=0$
if $p_i$ is not in the set $S$ and is $\chi_i=1$
if $p_i$ is in the set $S$,
where ${\bf\chi}$ is the characteristic vector
with $n$ components $\chi_i$,
where ${\bf f}$ is the probability vector with $n$ components $f_i$,
and where $\cdot$ denotes the dot product with the Euclidean metric
in $n$ dimensions.

	One can combine sets of perceptions by the AND ($\wedge$)
and OR ($\vee$) operations.  For example, if $S$ and $S'$ are two sets,
with characteristic functions $\chi_i$ and $\chi'_i$ respectively,
then the set $\hat{S} = S\wedge S'$
has the members common to both $S$ and $S'$
and hence the characteristic function $\hat{\chi}_i = \chi_i\chi'_i$,
and the set $\check{S} = S\vee S'$ has the members that occur
in either $S$ or $S'$ (or in both) and hence the characteristic function
$\check{\chi}_i = \chi_i+\chi'_i-\chi_i\chi'_i$.
Then, for example,
 \begin{equation}
 f(\hat{S}) \equiv f(S\wedge S') = \hat{\bf\chi}\cdot{\bf f}
  = \sum_{i=1}^n \chi_i\chi'_i f_i.
 \label{2}
 \end{equation}
The vector $\hat{\bf{\chi}}$ thus has the components
$\hat{\chi}_i = \chi_i\chi'_i$ and so is determined by the
vectors $\bf{\chi}$ and $\bf{\chi'}$, but the determination
is not simply a usual product of the vectors
as it is for the components.

	This awkwardness in expressing
$\hat{\bf{\chi}}(\bf{\chi},\bf{\chi'})$
can be taken as a motivation to replace the characteristic vectors
by diagonal projection matrices,
whose diagonal elements are the components
of the characteristic vectors.
E.g., let the projection matrix for the set $S$ be
 \begin{equation}
 P(S) = \mathrm{diag}(\chi_1,\chi_2,\ldots,\chi_n),
 \label{3}
 \end{equation}
with components
 \begin{equation}
 P_{ij}(S) = \chi_i \delta_{ij},
 \label{4}
 \end{equation}
where $\delta_{ij}$ is the Kronecker delta symbol that is
0 if $i \neq j$ and 1 if $i=j$,
and where the Einstein summation convention is not being used.
Then one can express the projection matrix for $S\wedge S'$ simply as
 \begin{equation}
 P(S\wedge S') = P(S) P(S').
 \label{5}
 \end{equation}
The simplicity of Eq. (\ref{5}) will be taken as the main motivation
for going to the matrix form of the characteristic values.

	Now to express the probability of a set $S$,
one also needs to replace the probability vector $\bf{f}$
by a diagonal probability matrix or density matrix
 \begin{equation}
 \rho = \mathrm{diag}(f_1,f_2,\ldots,f_n),
 \label{6}
 \end{equation}
with components
 \begin{equation}
 \rho_{ij}(S) = f_i \delta_{ij},
 \label{7}
 \end{equation}
again with no sum over $i$.
 
	One may get this density matrix directly from
the imperceptible time evolution of the dynamical system
if we define the time-dependent matrix
 \begin{equation}
 R(t) = \mathrm{diag}(r_1(t),r_2(t),\ldots,r_n(t)),
 \label{7b}
 \end{equation}
where $r_j(t)=0$ during the time that $i(t)\neq j$,
and $r_j(t)=1$ during the time that $i(t) = j$.
In other words, at each time $t$, $R$ is the matrix with all 0 entries,
except for a single 1 on the diagonal at the position of $i(t)$,
the state of the dynamical system at the time $t$.
Thus the components of $R(t)$ are
 \begin{equation}
 R_{ij}(t) = \delta_{ij}\delta_{ji(t)}.
 \label{7c}
 \end{equation}
Then the density matrix $\rho$ is simply the time average of $R(t)$
(over an integer number of periods $T$ or else over an infinite time).

	Although $i(t)$, and hence also $R(t)$, which just encodes
the information in $i(t)$ in a different form,
is not directly relevant to the perceiver,
which has no access to the time $t$,
the time-average of $R(t)$, namely $\rho$,
is relevant to the perceiver by giving the measure
of the various perceptions that the perceiver has.
Thus we can say that it is $\rho$ that encodes the properties
of the dynamical system that are relevant for the perceiver.
If we simply call these the relevant properties of the dynamical system,
we can, if we desire, dispense with the perceiver that was invoked
to motivate the distinction between ``relevant'' time averages
and the ``irrelevant'' detailed time dependence of the dynamical system.

	Having thus defined the density matrix $\rho$
of the classical system to give the fractions of time
that the system spends in its various states,
the only ``relevant'' properties of the state of the dynamical system,
we can see that the probability that a randomly-selected perception
is in the set $S$ is simply
 \begin{equation}
 f(S) = \mathrm{tr}(P(S) \rho)
  \equiv \sum_{i=1}^n \sum_{j=1}^n P_{ij}(S) \rho_{ji}.
 \label{8}
 \end{equation}

	Thus we have expressed the hypothetical
probabilities (actually fractions of an unperceived time),
for a set $S$ of perceptions in this classical model,
by Eq. (\ref{8}), the trace of the product
of a diagonal projection matrix and a diagonal density matrix.

\section{Generalization to Quantum Mechanics}

\hspace{.25in}Now it is straightforward to generalize
this algorithm for the probabilities in a classical model
to obtain the rule for the probabilities in a quantum model:
Simply let $P(S)$ be replaced by any Hermitian projection matrix,
not necessarily diagonal or real,
and let $\rho$ be replaced by any positive Hermitian matrix
with unit trace, also not necessarily diagonal or real.
(The trace condition is so that one gets unit probability
for the complete set of perceptions, which is represented
by the projection matrix that is the unit matrix.)
Then $\rho$ represents the density matrix of the quantum system
(itself a projection matrix if the quantum state is pure,
though there is no need to restrict to this special case),
$P(S)$ is the matrix representation of the projection operator
onto a set of orthogonal states that corresponds to
a set $S$ of perceptions, 
and Eq. (\ref{8}) gives the probability for that set of perceptions
in the given quantum state.

	The right hand side of Eq. (\ref{8}) is invariant
under a unitary transformation of both the projection
operator and the density matrix,
 \begin{equation}
 \mathrm{tr}(P(S) \rho) = \mathrm{tr}(\tilde{P}(S) \tilde{\rho}),
 \label{9}
 \end{equation}
where
 \begin{equation}
 \tilde{P}(S) = U P(S) U^{\dagger}, \ \tilde{\rho} = U \rho U^{\dagger}, 
 \label{10}
 \end{equation}
with $U$ being a unitary matrix, $U^{\dagger} = U^{-1}$.
Applying this unitary transformation is equivalent to
choosing a new basis of pure states for the quantum system.

	If one chooses the unitary transformation so that
$\tilde{P}(S)$ is diagonal, then in the corresponding basis
$\tilde{P}(S)$ is the projection operator onto a subset of these basis
states, the subset being those pure states that are eigenvectors
of $\tilde{P}(S)$ with unit eigenvalue.
In this basis the projection operator looks classical,
but if in this basis the density matrix $\tilde{\rho}$ is not diagonal,
then the quantum state does not look classical.

	Conversely, if one chooses the unitary transformation so that
$\tilde{\rho}$ is diagonal, then the elements of $\tilde{\rho}$
represent the probabilities for the corresponding pure basis states.
In this basis the quantum state looks classical,
but if in this basis the projection operator
$\tilde{P}(S)$ is not diagonal, then it is not a classical
projection operator simply onto a subset of these basis states.

	If we consider two generic projection operators, say
$P(S)$ and $P(S')$, that cannot be diagonalized in the same basis
and hence do not commute, then $P(S) P(S')$ will not be
a projection operator, so in the generic quantum case we lose
Eq. (\ref{5}) that was our original motivation (in the classical case)
for considering the projectors as matrices.
Hence the procedure used above, to get to quantum mechanics and
its Eq. (\ref{8}) for probabilities from classical mechanics,
is somewhat paradoxical in that the motivation that led to Eq. (\ref{8})
in the classical case no longer applies in the quantum case.

	But this situation is not without parallel in previous
physical arguments.  For example, there was the motivation,
from knowledge of the medium (with a preferred rest frame)
for sound waves,
to consider an ether (with its postulated preferred rest frame)
as a medium that could oscillate to give electromagnetic waves.
But then once Maxwell's equations for these latter waves were known,
one could show (admittedly, after several decades had elapsed)
that the equations have the symmetry of the Lorentz transformations,
so there is no effect of any postulated preferred rest frame
for an ether.  Thus although an ether may have had
some motivational value in leading to Maxwell's equations
for electromagnetic waves, once these equations were found
and understood, one lost the motivation for an ether with
a preferred rest frame.

	Somewhat analogously, here Eq. (\ref{5}) was the motivation
from an analysis of classical mechanics for getting Eq. (\ref{8}),
but after Eq. (\ref{8}) was generalized from the real diagonal
matrices applicable to classical mechanics to the more general
Hermitian matrices that give quantum mechanics,
one loses Eq. (\ref{5}).

	Just as we argued above that the real diagonal $n\times n$
matrix $\rho$ for a classical system with $n$ states
represented the full relevant properties of the deterministic
dynamical state of that system,
so we can postulate that the Hermitian $n\times n$
matrix $\rho$ for a quantum system with $n$ basis states
represents the full relevant properties of the dynamical state
of that system.  Thus one can postulate that the relevant properties
of a quantum system are completely described by its density matrix,
with no fundamental uncertainty or probabilities.
It is only when one hypothesizes a random sample
of a set $S$ of perceptions, here represented by the projection
operator $P(S)$, that one gets by Eq. (\ref{8}) the probability
$f(S) = \mathrm{tr}(P(S) \rho)$, the measure of the set $S$
represented by $P(S)$.

\section{Superselection rules in quantum mechanics}

\hspace{.25in}If the quantum system and the perceptions of it
are restricted by superselection rules for some conserved quantity,
then in a basis of eigenstates of this quantity,
the relevant density matrix $\rho$ is restricted to have
all off-diagonal terms zero that connect different values of the
conserved quantity.
In other words, $\rho$ must be block-diagonal,
with each component within a block representing the same value
of the conserved quantity.

	For example, consider the case in which the quantum system
has a time $t$ and an evolution with respect to this time that is given
by a time-independent Hamiltonian,
but assume the perceiver has no access to this time.
From a viewpoint in which the time can be seen,
the quantum system can have a time-dependent density matrix.
However, the perceiver has no access to this,
and therefore what is relevant to it is only
the time average of the time-dependent density matrix,
which is the time-independent density matrix $\rho$
given above.

	One can then easily see that if one goes to a basis
of eigenstates of the Hamiltonian (energy eigenstates),
the time-independent density matrix $\rho$ must be block-diagonal,
having no off-diagonal terms connecting eigenstates of different
energies, since these terms would have an oscillatory time dependence
that is washed out whan one does the time average to get
the time-independent density matrix $\rho$.

	Thus when there are superselection rules,
the relevant density matrix $\rho$ of a quantum system
cannot be a generic positive Hermitian matrix of unit trace
but has further superselection restrictions.

	Alternatively, if one allows the quantum state density matrix
$\rho$ to have ``irrelevant'' properties, such as a time-dependence,
one can restrict to purely relevant probabilities $f(S)$
in Eq. (\ref{8}) in the presence of superselection rules
by restricting the projection operators $P(S)$ to obey the
superselection rules.  For example, in the case above
in which the time $t$ is imperceptible,
one can restrict each $P(S)$ to be block-diagonal in a basis
of eigenstates of the Hamiltonian (i.e., in the energy representation).
Then the probabilities will not depend upon the irrelevant properties,
even if the density matrix is allowed to depend upon them.

	However, for a discussion of the purely relevant properties
of the system and of the perceptions of it,
it may be cleaner to restrict both
the density matrix and the projection operators by the superselection
rules so that they do not have irrelevant properties.

\section{Generalizations other than ordinary quantum mechanics}

\hspace{.25in}We have seen that Eq. (\ref{8}) represents
a probability, under a certain random sampling,
in classical mechanics when one has diagonal projection
operators and diagonal density matrices,
and that one can generalize this to quantum mechanics
by allowing generic Hermitian projection operators
and generic positive unit-trace Hermitian density matrices,
subject only to superselection constraints.
However, there are other ways in which one might consider
generalizing the quantities in Eq. (\ref{8}).

	For example, one might generalize the real diagonal matrices
to nondiagonal matrices that are still real and othogonal rather
than being Hermitian.  This would then give real quantum mechanics
rather than ordinary complex quantum mechanics,
a restriction on quantum mechanics that would be an intermediate
case between classical mechanics and ordinary quantum mechanics.

	A generalization that would be more general than
ordinary quantum mechanics as given above would be
simply to require that $P(S)$ and $\rho$ be positive Hermitian
matrices, but not requiring that $P(S)$ be a projection operator
or that $\rho$ have unit trace.
Then $f(S) = \mathrm{tr}(P(S) \rho)$ given by Eq. (\ref{8})
would no longer directly be a probability
but could be reinterpreted to be the measure
for the set $S$ of perceptions.
This generalization is what I have elsewhere called
Sensible Quantum Mechanics
\cite{SQM,SQM2,MS}.

	With the positive operator $P(S)$ no longer being restricted
to be a projection operator,
one could define it to have the additivity property
that for two disjoint sets $S$ and $S'$,
the positive operator for the union of the sets
is the sum of the positive operators for the individual disjoint sets,
 \begin{equation}
 P(S\vee S') = P(S) + P(S') \ \mathrm{if} \ S\wedge S' = 0, 
 \label{11}
 \end{equation}
the null set.
With this property and an analogous one that would apply
in the case of infinite unions, the positive operator $P(S)$
would become a positive-operator-valued (POV) function
\cite{Dav}.

	If the set of all possible perceptions is $M$,
and if $f(M) = \mathrm{tr}(P(M) \rho)$ is finite
(a nontrivial restriction),
then one could take $f(S)/f(M)$ to be the probability
of the set $S$, the probability that a perception
in that set would be chosen if one made a random sample
from the set of all possible perceptions with measure
given by $f(S')$ for each set $S'$.
Thus in this case one can say that a density matrix $\rho$
for the quantum system induces, by the normalized
form of Eq. (\ref{8}), a probability distribution
over the space of sets of perceptions.

	If $f(M)$ is not finite, one cannot define
a normalized probability distribution over the full set $M$
of perceptions, but one can still do it over some
subset, say $M'$, if $f(M')$ is finite.
Then one could calculate the conditional probability
$f(S')/f(M')$ that a perception is in the subset $S'$
of $M'$, given that it is in the normalizable set $M'$.

\section*{Acknowledgments}

\hspace{.25in}I am grateful for lectures by, and discussions with,
Lucien Hardy and Anton Zeilinger at the Science \& Ultimate Reality
meeting, which motivated my consideration of this subject.
I am also grateful to the hospitality of Flushing Meadows Corona Park,
Queens, New York, for providing diversions for my daughters
while I formulated the basic ideas for this paper.
Financial support has been provided by the Natural Sciences and
Engineering Research Council of Canada.


\end{document}